\documentclass[11pt,a4paper]{article}
\pdfoutput=1
                   
\usepackage{jheppub}
\usepackage{multirow}
\usepackage{color}
\usepackage[usenames,dvipsnames,table]{xcolor}
\usepackage{graphicx}
\usepackage{epsfig}  
\usepackage{epsf}    
\usepackage{dcolumn}
\usepackage{bm}
\usepackage{dcolumn}
\usepackage{textcomp}
\usepackage{float}
\usepackage{subfig}
\usepackage{hypcap}   
\usepackage{slashed}
\usepackage{tablefootnote}
\usepackage[normalem]{ulem}
\usepackage{amsmath}
\newcommand{\stkout}[1]{\ifmmode\text{\sout{\ensuremath{#1}}}\else\sout{#1}\fi}
\usepackage[]{hyperref}
  
\usepackage{color}
\usepackage{pifont}
\usepackage{array,multirow}
\usepackage{url}
\hypersetup{
  bookmarks=true,         
  unicode=false,          
  pdftoolbar=true,        
 pdfmenubar=true,        
 pdffitwindow=true,     
 pdfstartview={FitH},    
 pdfsubject={Extra Dimensional Model},   
 pdfnewwindow=true,      
 pdfcreator={RevTeX},
 colorlinks=true,       
 linkcolor=red,          
 citecolor=blue,        
 filecolor=black,      
 urlcolor=blue,           
  }

\newcommand{\g}{\gamma}
\newcommand{\MD}{M_D}
\newcommand{\Rinv}{R^{-1}}
\newcommand{\muhat}{\hat{\mu}}
\newcommand{\nuhat}{\hat{\nu}}
\newcommand{\vecn}{\vec{n}}

\makeatletter
\renewcommand{\fnum@table}{\textbf{\tablename~\thetable}}
\renewcommand{\fnum@figure}{\textbf{\figurename~\thefigure}}
\makeatother

\title{'Fat-brane' Universal Extra Dimension model confronted with the ATLAS multi-jet and photonic searches at 13 TeV LHC}

\author[a]{Esra Akyumuk}
\author[b]{Durmus Karabacak}

\affiliation[a]{Department of Physics, Middle East Technical University, TR06800 Ankara, Turkey}
\affiliation[b]{Department of Energy System's Engineering, Faculty of Technology, Mugla Sitki Kocman University, TR48000 Mugla, Turkey}

\emailAdd{esra.akyumuk@metu.edu.tr}
\emailAdd{durmuskarabacak@gmail.com}

\abstract
    {The current status of `fat-brane' minimal Universal Extra Dimensions (fat-mUED) is studied in the light of ATLAS experiment's recent reports. At the Large Hadron Collider (LHC) color charged first level Kaluza-Klein (KK) particles (first level excited quarks and gluons) can be abundantly pair-produced due to conserved quantity, {\it viz.}, KK-parity, and strong interaction. The cascade decay of these particles to one or more Standard Model (SM) particle(s) and lighter first level KK particle(s) stops after producing the lightest excited massive state, named as the lightest KK particle (LKP). With the presence of gravity induced decays, stability of the LKP is lost and it may decay to photon or Z-boson by radiating KK-excited gravitons, hence leading to final state with photon(s) at the LHC. A variant signal topology is established when pair-produced first level colored KK particles undergo direct decay to an associated SM partner along with KK-excitations of graviton; thus leading to a signal with two hard jets and substantial missing energy. The ATLAS experiment lately reported two searches at $13$ TeV LHC with $139$ inverse-femtobarn of data; (i) multi-jet and (ii) photon and jets with missing energy. In both searches, the results showed no substantial deviation from the number of background events of the SM. Provided the absence of any number of excess events in both searches we constrained the parameters of the fat-mUED model, {\it viz.}, the higher-dimensional Planck mass and the compactification scale.           
}
    

\begin{document}
\maketitle
\flushbottom

\section{Introduction}\label{intro}
 
The Large Hadron Collider (LHC) experiment, already collected $139$ fb$^{-1}$ amount of data, is in phase of Run-III as of now, aims to reveal any new physics (NP) scenario beyond the Standard Model (SM). In terms of new degrees of freedom, so far the only particle, the Higgs boson with its mass at the order of electroweak (EW) scale and with its high degree of resemblance to the SM Higgs \cite{ATLAS:2019nkf, CMS:2018uag}, has been revealed at the LHC. This, by itself, is another milestone in the history of particle physics, although the SM still comprises shortcomings on theoretical considerations such as the Higgs mass/hierarchy problem. In addition, the SM does not contain a new weakly interacting massive particle, the dark matter (DM), needed to explain the bulk mass of the Universe. Another notable issue with the SM is that it does not incorporate gravity in its framework as other fundamental forces. These shortcomings clearly point NP beyond the SM (BSM).

Some (or all) of the aforementioned issues of the SM can be mitigated through postulating extra spatial dimensions besides our usual ($3+1$) dimensional spacetime. The Arkani-Hamed, Dimopoulos, Dvali (ADD) \cite{Antoniadis:1998ig, Arkani-Hamed:1998sfv} model, a candidate for addressing the hierarchy/naturalness problem for instance, assumes localization of the SM fields to a $3$-brane embedded in a higher dimensional bulk. Only gravity is presumed to percolate into $D=(4+\delta)$ dimensional bulk with $\delta$ number of {\it large} extra dimensions of size at the order of $\sim$ eV$^{-1}-$ keV$^{-1}$, for values in the range of $\delta=2-6$. This construction then results in dilution of ($3+1$) dimensional Planck mass $M_{Pl}$, explaining the hierarchy between the EW and the Planck scale, and hence nullifying the problem.      
The very same problem is also approached in Randall-Sundrum (RS) \cite{Randall:1999ee, Randall:1999vf} model which posits the existence of a single additional spatial dimension, yet the geometry is characterized by the anti-de Sitter metric as opposed to the flat metric. Using this particular geometric construction, the gravity percolating in the fifth dimension experiences an exponential decrease. In the plethora of models with extra spatial dimension(s), there is a class of models, wherein SM particles (partially or fully) can access to {\it universal} and {\it small} ($\sim$ TeV$^{-1}$-sized) extra dimension(s) \cite{Antoniadis:1990ew, Appelquist:2000nn, Cheng:2002ab}. Apart from offering a diverse range of possibilities for collider physics \cite{Appelquist:2000nn, Cheng:2002ab, Belyaev:2012ai, Belanger:2012mc, Ghosh:2008ix, Ghosh:2008dp, Ghosh:2018mck, Beuria:2017jez, Dey:2014ana, Edelhauser:2013lia, Flacke:2012ke, Nishiwaki:2011gm, Murayama:2011hj, Choudhury:2011jk, Ghosh:2010tp, Bhattacherjee:2010vm, Bertone:2010ww, Freitas:2007rh, Ghosh:2012zc, Ghosh:2008ji, Choudhury:2009kz, Rizzo:2001sd, Muck:2003kx, Bhattacharyya:2005vm, Battaglia:2005zf, Bhattacherjee:2005qe, Datta:2005zs, Datta:2005vx, Bhattacherjee:2007wy, Bandyopadhyay:2009gd, Datta:2011vg, Cheng:2002rn, CMS:2011bsw, CMS:2018dqv, CMS:2011xuw, CMS:2014jvv, ATLAS:2015yln, ATLAS:2015shg, Cacciapaglia:2013wha, Choudhury:2016tff, ATLAS:2015rul, ATLAS:2015hef}, Universal Extra Dimensions (UED) models may potentially introduce a new mechanism for Supersymmetry (SUSY) breaking \cite{Antoniadis:1990ew}, decrease the unification scale to a few TeVs \cite{Dienes:1998vg, Dienes:1998vh, Hossenfelder:2004up, Bhattacharyya:2006ym}, alleviate the upper bound on the mass of the lightest Higgs boson of Supersymmetry (SUSY) \cite{Bhattacharyya:2007te}, construe the Higgs boson as a composite particle of quarks, and thereby nullifying the need of a fundamental scalar field \cite{Arkani-Hamed:2000ulw}, furnish a viable candidate for dark matter (DM) in the universe \cite{Servant:2002aq, Cheng:2002ej, Kong:2005hn, Hooper:2007qk, Flacke:2017xsv, Belanger:2010yx, Kakizaki:2006dz, Burnell:2005hm, Ishigure:2016kxp, Cornell:2014jza, Dobrescu:2007ec}. In particular, UED models with two extra spatial dimensions (2UED) \cite{Dobrescu:2004zi, Burdman:2005sr} possess compelling characteristics in addition to the generic benefits associated with the UED framework. Specifically, the 2UED model proffers a natural explanation for the longevity of proton decay \cite{Appelquist:2001mj}, and intriguingly, predicts that fermions should come in three generations \cite{Dobrescu:2001ae}.

In the minimal UED (mUED) model there exists, while maintaining the gauge group of the SM as it is, a single inverse-TeV sized extra dimension which is then compactified on a half circle $S_1/Z_2$ of radius $R$. All fields in the SM are then permitted to access this universal extra dimension. This kind of spacetime construction in UED models results in a tower of infinitely many new degrees of freedom, collectively called as the Kaluza-Klein (KK) excitations with masses $\sim n\Rinv$, where $n$ is an integer called as the KK number signifying the associated excitation level. Then one assigns $n=0$ to the SM fields and $n>0$ to their higher level KK excitations. An attractive characteristics of UED models come from the conservation of momentum for a particle moving along the extra dimension. This conservation in higher dimensional model then leads to a selection rule \textit{viz.}, the KK-number conservation, in ($3+1$) dimensions. The KK-number is broken at $1$-loop level and leaves behind a conserved quantity called as the KK-parity ($\equiv (-1)^n$). Similar to the discrete symmetry in SUSY models \textit{viz.}, the R-parity, the preservation of the KK-parity within UED models guarantees that the first level KK particles are produced exclusively in pairs in high energy particle collider experiments, and the lowest massive state in the KK spectrum, referred to as the Lightest Kaluza-Klein particle (LKP), does not decay assuming it is neutral. It also grants an appealing feature to UED models that the constraints from EW precision data are rather weak. 

One of the possible modification of the above scenario may come from embedding spacetime structure of UED into $D=(4+\delta)$-dimensional spacetime (the bulk) to which both SM particles and gravity may percolate. In this setup, a TeV-sized Planck mass can be achieved with large (of order eV$^{-1}$ to keV$^{-1}$-sized) extra dimensions as in the ADD. However, this symmetric configuration faces with the fact that ordinary SM fields would also have eV-massed KK excitations, and therefore, contradicts with the null results of collider experiments. A way of circumventing this problem may be to contemplate an asymmetric configuration wherein spread of the SM fields is restricted to $(3+m)$-brane with $m$ small (of order TeV$^{-1}$) extra dimensions embedded into $D=(4+\delta)$-dimensional bulk with $\delta$ number of large extra dimensions ($\sim$ \text{eV}$^{-1}$ to \text{keV}$^{-1}$\text{-sized}) to which only gravity is unreservedly permitted to percolate. Postulating that one of the compactified small extra dimension(s) is aligned with one of the large extra dimension(s) results in $\sim \mathcal{O}(1)$ TeV Planck scale in $4$D, and also evades existing collider constraints on eV-massed excitations. In this construction, the matter fields are confined to small but non-vanishing width (or extension) of $(3+m)$-brane in the bulk to which only the gravity can percolate. This configuration presents a rich landscape for potential signals that can be sought at current or future high-energy particle colliders, such as the LHC. This kind of realization of UED model is called as the `fat-brane'-UED scenario \cite{DeRujula:2000he, Dicus:2000hm, Macesanu:2002db, Macesanu:2002ew, Macesanu:2005wj, Macesanu:2003jx} and the LHC phenomenology of it is the main concern of the present study.           
    
In the case of fat-brane UED models, the KK-parity is broken because of the interactions induced by the gravity. It allows the first level KK particles undergo direct decay into their corresponding SM fields through emission of KK gravity excitations ($X^{n=1} \rightarrow X + \hat{G}$ where $X(X^{n=1})$ is the SM particle (the first level KK excitation) and $\hat{G}$ representing the KK graviton. Moreover, the LKP can now decay gravitationally, and hence, is no longer a stable particle as in the case of UED. Decay of LKP into $\g$ or $Z$-boson makes the collider physics of `fat-brane' UED models quite interesting. In addition to their direct decay induced by the gravity, decay cascade of the first level KK particles may continue with the KK-number preserving decays (KKPD), which involve decay to lighter first level KK particles along with radiating SM particles. This cascade of decays continues until reaching to the least-massive KK particle. Subsequently, the LKP undergoes further decay through gravity mediation, resulting in the emission of a photon ($\gamma$) or a $Z$-boson. The production cross-section of the first level colored particles (first level quarks and gluons) will be substantially larger than that of excited electroweak gauge bosons because of their color charges and strong coupling. The phenomenology of such particles will be mainly determined by the relative strength of two competing decay mechanisms, namely gravity induced decays (GID) and KKPD. If the former decay mechanism the GID is larger than the KKPD, then the decay of pair-produced first level colored KK particles would be characterized by two hard jets plus large missing energy due to gravity excitations escaped from detection. Contrarily, if the KKPD surpasses the GID, then the first level quarks and/or gluons cascade decay to lighter first level KK states and a SM particle. The cascade terminates with the decay of LKP into either $\g$ or $Z$-boson gravitationally. In such a situation, the final state would be characterized by $\g \g, ZZ \text{~or~} \g Z + X + \slashed{E}_{T}$ wherein X refers to jets and/or leptons radiated in the decay cascade. Alternatively, it is possible that one of the partons decays through the KKPD chain and the decay of other parton is mediated by the GID. For this case, the final state signal would be comprised of  jets, $\g\text{~or~} Z$ and $\slashed{E}_{T}$ carried off by the graviton excitations. 

The ATLAS Collaboration lately published the results of multi-jet \cite{ATLAS:2020syg} and $\text{jets} + \g$ plus $\slashed{E}_{T}$ \cite{ATLAS:2022ckd} final state searches at $\sqrt{s} = 13$ TeV LHC with $139$ fb$^{-1}$ of integrated luminosity data. In these searches, the ATLAS Collaboration was seeking traces of squarks and gluinos (Supersymmetric partners of quarks and gluons) in the LHC data. In both searches, the results failed to reveal any observable deviation from the expected number of background events as predicted by the SM. Hence, upper limits on the visible cross-sections as well as lower limits on the masses of such particles, in the case of simplified SUSY models, are set at $\%95$ confidence level (C.L.) in each search. The use of limits obtained in these searches is not restricted to SUSY models as similar final state topologies can also be realized in other NP models. In this letter, we confront the minimal fat-brane UED  (the fat-mUED) model with the null results of aforementioned ATLAS searches and constrain the parameter space of the model accordingly.  

Subsequent sections of this Letter are organized as the following. In the Section \ref{Model:mUED} we review in brief the minimal version of Universal Extra Dimensions (mUED) and in Section \ref{Model:fat-mUED} present the details of fat-mUED model. We discuss the LHC phenomenology of fat-mUED in Section \ref{sec:Collider_Phenomenology} and the exclusion limits on the model from ATLAS searches in Section \ref{sec:Exclusions}. In Section \ref{sec:Conclusion} we present Summary and Conclusions. 

\section{The minimal UED}\label{Model:mUED}
The minimal UED (mUED) model is constructed via single small ($\mathcal{O} \sim$ TeV$^{-1}$) size flat spatial extra dimension (denoted by $y$). Its compactification on a half-circle (with radius of $R$) is then achieved via $Z_2$ orbifolding. All SM fields are permitted to access into extra dimension. An additional $Z_2$ symmetry, identifying $y \rightarrow -y$, is necessary for obtaining chiral fermionic structure of the SM in $4$D and for removing unwanted degrees of freedom in gauge fields. It creates fixed boundary points located at $y=0$ and $y=\pi R$ and breaks the translational symmetry along $y$. Each component of the field in higher dimensions exhibits parity, being either even or odd. Low energy Lagrangian obtained after compactification comprises of zero-mode fields, which are congruent via the SM fields, and the higher mode KK excitations forming a tower with increasing mass. The details of KK decomposition can be found at Ref.~\cite{Appelquist:2000nn, Cheng:2002ab} for interested readers.

In the mUED model~\cite{Antoniadis:1990ew, Appelquist:2000nn, Cheng:2002ab}, fermions, gauge bosons and the Higgs are granted access to percolate through the compactified small extra dimension, and the momentum along $y$ is both conserved and quantized. The momentum conservation in higher dimensional theory is then subsequently expressed as preservation of the KK-number ($n$) in $4$D effective theory. On the other hand, the fixed boundary points at $y=0, \pi R$ break the translational invariance along $y$ and therefore the KK-number should not be treated as a good quantum number. Nevertheless, if local operators at fixed points are chosen such that they are symmetric under $y \leftrightarrow -y$, then a remnant conserved quantity named as the KK-parity $(\equiv (-1)^n)$ stays intact. The existence of this discrete symmetry has significant implications, as it precludes the production of the first level KK particles singly at collider experiments and results in the stability of the lightest KK particle. In the case of mUED the lowest lying massive state corresponds to the first level KK excitation of $B_{\mu}^{1}$, the gauge field of $U(1)_{Y}$. Additionally, tree-level exchange of the KK modes does not contribute to quantities that can currently be measured. Consequently, the corrections to EW observables are suppressed at the loop level. The mass of $n$-th KK mode of excitation is determined via $m_{X^{(n)}}^2 = m_{X}^2+(n \Rinv)^2$, wherein $m_{X}$ is the mass term belonging to the corresponding SM field. This relation manifestly shows that the mUED particle spectrum at any KK level of $n$ is quite degenerate and the splitting between different KK particles of the same KK level may only be significant regarding top quark and the Higgs boson cases. With this mass spectrum picture, most of the first level KK particles would be stable. Fortunately, the compressed mass spectrum at tree-level could be somewhat relaxed through inclusion of two distinct types of loop corrections~\cite{Cheng:2002iz} and prompt decay of KK excitations can be achieved. The bulk corrections that result from loop diagrams around the compactified dimension, are non-zero and finite only for the KK gauge bosons. Violation of $5$D Lorentz and translational invariance at orbifold boundaries produce additional corrections. These corrections are proportional to log$(\Lambda^2)$, wherein $\Lambda > \Rinv$ is the cut-off scale up to which the theory is valid.\\
The mass-squared matrix in the basis of ($B_{\mu}^{n}, W_{\mu}^{3(n)}$), when diagonalized gives the mass eigenstates and eigenvalues corresponding to (physical) photon and $Z$-boson, is the following: 
\begin{equation*}
\mathcal{M}^2_{B_\mu^{n},W_\mu^{3(n)}}=
\begin{pmatrix}
(n\Rinv)^2 + \hat{\delta}m_{B^{n}}^2 +\frac{1}{4}g_1^2v^2 & \frac{1}{4}g_1 g_2 v^2 \\
 \frac{1}{4}g_1 g_2 v^2 &  (n \Rinv)^2 + \hat{\delta}m_{W^n}^2+\frac{1}{4}g_2^2v^2  
\end{pmatrix}
,
\end{equation*}
in which, $\hat{\delta}m_{B^{n}}^2$ and $\hat{\delta}m_{W^n}^2$ are the total (boundary and bulk) $1$-loop corrections for $B_\mu^n$ and $W_\mu^{3(n)}$, $g_1$ and $g_2$ are $U(1)_{Y}$ and $SU(2)_{L}$ gauge couplings, and $v=246$ GeV is the Higgs vev., correspondingly. We note that even for $\Rinv \ge 500$ GeV, the mixing is small enough and becomes progressively smaller for the higher KK levels. Therefore, the mass eigenstates of neutral (physical) KK gauge bosons ($\g^1, Z^1$) are in align with the corresponding gauge eigenstates $B_{\mu}^1, W_{3 \mu}^{1}$, in practice. The LKP of mUED, has no electric charge or color, is weakly interacting and massive. These attributes render it a viable candidate for the cold DM. The mass of $\g^{1}$, depending chiefly on $\Rinv$ and $\Lambda$ (though logarithmically), can determine the DM relic density in the universe. Measurements from WMAP-Planck \cite{WMAP:2010sfg, Planck:2015fie} establish an upper bound of $1/R < 1400$ GeV \cite{Hooper:2007qk} on the compactification scale of the mUED.

The colored first level KK states may be abundantly produced in pairs at the LHC. The quantum corrections, creating mass splitting among first level KK particles, allow their subsequent decay. Therefore, the pair-produced strongly interacting first level excited particles (excited gluons and quarks), with their subsequent decay to lighter first level KK particles and an associated SM particle, results in multi-jet, multi-lepton plus missing energy signal coming from the LKP escaping from detection at the LHC. Given the upper bound on $\Rinv$ from DM relic density results and the parameters of the model studied in Ref.\cite{Avnish:2020atn} with the ATLAS $139$ fb$^{-1}$ multi-jet results at $13$ TeV LHC suggest that the minimal version of UED (mUED) is entirely ruled out. However, its non-minimal version \cite{Flacke:2013pla}, the mUED extended with brane-localized and fermion bulk mass terms, still survives.     
\section{The fat-mUED: Model description}\label{Model:fat-mUED}
After discussing the mUED model briefly, in this section we would like to present the details of a variant of UED models called as 'fat-UED'. The fat-UED scenario is realized through the incorporation of UED in a ($4+\delta$)D bulk, wherein $\delta$ represents the number of \textit{large} (of order inverse eV to keV) extra dimensions. In this framework, it is presumed that both the fields of SM and gravity are able to spread to the small extra dimension(s) of UED; however, access to the ($\delta - 1$) large extra dimension(s) is granted exclusively to the gravity. In this type of construction, the small extra dimension(s) lie along one of the large extra dimensions, and may be regarded as the width of matter brane in a ($4+\delta$)D bulk. It is worth highlighting that one may contemplate a scenario wherein all SM particles and gravity are granted unrestricted access to the complete bulk. However, given the non-observation of eV-massed KK particles at current collider experiments one is compelled to consider an asymmetric case which results in $m_{KK}\sim \mathcal{O}(1)$ TeV mass scale, and successfully evades experimental lower bounds. The inclusion of the fat-brane structure in UED brings another decay mode to the KK excitations, \textit{viz.} the gravity induced decays, and hence promises interesting collider signals not present in generic UED models.    

In this work, for the sake of simplicity, we assumed a single flat extra dimension, hence \textit{fat-mUED}, of size $R\sim$ TeV$^{-1}$ embedded in a bulk of dimensions $D=(4+\delta)$ wherein $\delta$ represents number of large (of order eV$^{-1}$ to keV$^{-1}$) extra dimensions. Compactifying $\delta$ number of large extra dimensions on a torus $T^{\delta}$ of $\delta$-dimensions of volume $V(\delta)\sim r^{\delta}$ establishes the following relation between the Planck mass in ($4+\delta$)D and the effective $4$D Planck mass $M_{Pl}$:  
\begin{equation}\label{eq:planck}
M_{Pl}^2 = M_{D}^{\delta+2}\bigg(\frac{r}{2\pi}\bigg)^{\delta}. 
\end{equation}
Denoting TeV$^{-1}$ size small extra dimension and eV$^{-1}$ size large extra dimensions that are accessible by the gravity by the coordinates $y=x^4$ and $x^5,x^6,x^7,....,x^{4+\delta}$ respectively, the interactions between the SM and the gravity fields in $D=(4+\delta)$-dimensional theory can be described by the action: 
\begin{equation}\label{eq:int_lagrangian}
  \mathcal{S}_{int}=\int d^{D}x \delta(x^5)...\delta(x^{D})\sqrt{-\hat{g}}~\mathcal{L}_{mat}, 
\end{equation}
 in which, $\mathcal{L}_{mat}$ is Lagrangian density for matter fields and $\hat{g}$ is $(4+\delta)$-dimensional linearized metric written as $\hat{g}_{\muhat \nuhat} = \hat{\eta}_{\muhat \nuhat} + \hat{\kappa} \hat{h}_{\muhat \nuhat}$ in which $\hat{\kappa}$ is given in terms of higher dimensional Newton's constant $G^{(4+\delta)}$ as $\hat{\kappa}^2 = 16 \pi G^{(4+\delta)}$. $\hat{h}_{\muhat \nuhat}$, a higher dimensional metric and a tensor in ($4+\delta$)D, contains the following parts: a $4$D tensor ($h_{\mu\nu}$ graviton), $\delta$ vectors ($A_{\mu i}$ gravi-photons) and $\delta^2$ scalars ($\phi_{ij}$ gravi-scalars):
\begin{equation}
  \hat{h}_{\muhat \nuhat} = V_{\delta}^{-1/2} 
  \begin{pmatrix}
    h_{\mu\nu}+\eta_{\mu\nu}\phi & A_{\mu i}\\
    A_{\nu j} & 2\phi_{ij}
   \end{pmatrix}
\end{equation}
in which $\mu, \nu = 0-3$, $i, j = 4-(3+\delta)$ and $\phi = \phi_{ii}$. Propagation into finite size large extra dimensions and compactification on a torus of $T^{\delta}$ results in the following KK expansion for the fields:
 \begin{equation}\label{eq:fields}
  \begin{aligned} 
      h_{\mu \nu} (x, y) &= \sum_{\vecn} h_{\mu \nu}^{\vecn}(x) \exp \bigg( \frac{ 2\pi i \vecn . \vec{y}}{r} \bigg )\\
      A_{\mu i} (x, y) &= \sum_{\vecn} A_{\mu i}^{\vecn}(x) \exp \bigg( \frac{2\pi i\vecn . \vec{y}}{r} \bigg )\\
      \phi_{i j} (x, y) &= \sum_{\vecn} \phi_{i j}^{\vecn}(x) \exp \bigg(\frac{2\pi i\vecn . \vec{y}}{r} \bigg) \\
  \end{aligned}
 \end{equation}
with $\vecn = \{n_1,...,n_{\delta} \}$. In Eq.~\ref{eq:fields}, $\vecn=\vec{0}$ and $\vecn>0$ correspondingly represent massless graviton, gravi-photons, gravi-scalars and a tower of massive KK-excitations. Masses of these higher level graviton, gravi-vector, and gravi-scalar, solely depending on KK-number $\vecn$ and the size of $r$, are degenerate and given by $m_{\vecn} = 2\pi|\vecn| r^{-1}$. The expression in Eq.~\ref{eq:int_lagrangian}, at the order of $\mathcal{O}(\hat{\kappa})$, reads as
\begin{equation}
  \mathcal{S}_{int} \supset -\frac{\hat{\kappa}}{2}\int d^{D=4+\delta}x ~ \delta(x^5)\delta(x^6) ... \delta(x^{4+\delta})\hat{h}^{\muhat \nuhat} T_{\muhat \nuhat},
\end{equation}
wherein, $T_{\muhat \nuhat}$ is given in $D=(4+\delta)$ dimensions by:
\begin{equation} \label{eq:em_tensor}
T_{\muhat \nuhat} = \bigg(-\hat{\eta}_{\muhat \nuhat} + 2\frac{\delta \mathcal{L}_{mat}}{\delta \hat{g}^{\muhat \nuhat}} \bigg)\Bigg|_{\hat{g}= \hat{\eta}}.
\end{equation}
The energy-momentum tensor, obtained for a given matter lagrangian density $\mathcal{L}_{mat}$ through above equation, then can be used to obtain gravity-matter interactions after expanding KK-modes and integrating over $x^4$ as:
\begin{equation}
  \begin{split}
    \mathcal{S}_{int} \supset & -\frac{\kappa}{2}\int d^4x\int_{0}^{\pi R}dy\sum_{\vecn}\bigg[ \bigg(h_{\mu \nu}^{\vecn} + \eta_{\mu \nu} \phi^{\vecn} \bigg)T^{\mu \nu}- 2 A_{\mu 4}^{\vecn}T_4^{\mu}+2\phi_{44}^{\vecn}T_{44}\bigg]\\
    & \times \exp \bigg(\frac{i 2\pi n_4 y}{r}\bigg),
    \end{split}
\end{equation}
wherein, $\kappa$ is the Newton's constant of $4$-dimensional theory and is given by $\kappa \equiv \sqrt{16 \pi G^{(4)}}=\hat{\kappa}/\sqrt{V_{\delta}}$.
The resulting Feynman rules for the interactions of matter and gravity, can be derived by using equations above, is quite complicated. We refer interested readers to Ref. \cite{Macesanu:2003jx, Macesanu:2005jx}. 
\subsection{The fat-mUED: Decay widths of first level KK excitations by Gravity Mediation}\label{sec:fat-mUED:GID}
After presenting the interactions of the gravity and the matter fields in the case of fat-mUED model above we are now ready to discuss the decay widths of the first level KK particles through gravity mediation with emphasis on the LHC experiment. The small extra dimension ($x^4=y$) of the fat-mUED model, accessible by the fields of both gravity and SM particles, is presumed to be the width of the brane along the large extra dimension(s) into that only the gravity can percolate. Such particular positioning of the SM-brane ($3$-brane) in $D=(4+\delta)$-dimensional bulk breaks translational invariance along $y$. Therefore, the conserved quantity of mUED (\textit{viz.}, the KK-parity) is no longer intact in the 'fat-brane' UED scenarios. This allows the first level KK excitations undergo direct decay to the associated SM particles, emitting gravity excitations (i.e. gravitons, gravi-photons, and gravi-scalars). The total decay width of first level matter fields through gravity mediation, then will be a sum of individual decay widths to gravitons, gravi-photons, and gravi-scalars with masses smaller than that of decaying particle:
\begin{equation}
  \Gamma  = \sum_{\vecn} \bigg( \Gamma_{h^{\vecn}} + \Gamma_{A^{\vecn}} + \Gamma_{\phi^{\vecn}} \bigg).
\end{equation}
The mass gap between different KK-modes of the graviton, which are chiefly determined by the compactification scale ($r$), is quite minuscule, given by $\Delta m = 2\pi/r \sim$ eV. This enables the aforementioned sum to be substituted with the following integral: 
\begin{equation}
  \sum_{\vecn} \Gamma_{\vecn} \Longrightarrow \int \Gamma_{\vecn} ~d^{\delta}\vecn,
\end{equation}
in which, $d^{\delta}\vecn$ corresponds to the number of $\vecn$-th modes of gravitons between $m_{\vecn}$ and $m_{\vecn}+dm$. As level-$\vecn$ KK graviton has mass of $m_{\vecn}^{2}=4\pi\vecn^2/r^2$, $\vecn^2$ is then $m_{\vecn}^2/\Delta m$. The number of KK gravitons in range of $(m_{\vecn}, m_{\vecn}+dm)$ can be found by using the volume of annular region between two  $D=\delta$ hyperspace with radii $m_{\vecn} / \Delta m$ and $(m_{\vecn}+dm)/\Delta m$:
\begin{equation}
  d^{\delta}_{\vecn} = \bigg(\frac{m_{\vecn}}{\Delta m}\bigg)^{\delta-1} \frac{dm}{\Delta m}d\Omega = \frac{m_{\vecn}^{\delta-1}}{\Delta m^{\delta}} dm ~d\Omega,
\end{equation}
wherein $d\Omega$ is the solid angle in dimension $D=\delta$. One may obtain $\Delta m^{\delta} = M_{D}^{\delta+2}/M_{Pl}$ relationship by using Eq.~\ref{eq:planck}, and then, find the total GID width of the first level KK particle by evaluating the expression below:
\begin{equation}
\Gamma = \frac{M_{Pl}^2}{M_{D}^{\delta+2}} \int \Gamma_{\vecn}^{\delta-1} dm~d\Omega.
\end{equation}
\section{LHC Phenomenology}\label{sec:Collider_Phenomenology}
After presenting the details of fat-mUED model in Section \ref{Model:fat-mUED}, and the tools necessary for the gravity induced decay (GID) width calculations in Section \ref{sec:fat-mUED:GID}, here we discuss in detail the collider physics of strongly produced first level KK particles particularly in the context of the LHC experiment. We restrict our discussion to strongly produced colored first level KK particles \textit{viz.}, the first level quarks and gluons, as their production cross-section is order of magnitude larger than that of color singlets. At the KK level of $n=1$ the excited particle content of the model considered here will be the same as that of the mUED; excited gauge bosons (gluon $g^{1}$, electroweak gauge bosons: ${W^{\pm 1}}$, $Z^{1}$, and photon $\g^{1}$), excited fermions (quark/lepton doublets of $SU(2)_L$: $Q^{1}/L^{1}$ and singlets $q^{1}/e^{1}$), and Higgses. The loop corrections, being proportional to logarithm of $\Lambda$, where $\Lambda$ is the cutoff scale above that some new dynamics take place, relax the degeneracy in the mass spectrum, and therefore allow the decay of first level KK particles. The numerical calculations suggest that the excited gluon ($g^{1}$) and the hypercharge gauge boson ($\g^{1}$) are the heaviest and the lightest excitations in the spectrum, respectively. At a given KK level of $n$, the cutoff scale $\Lambda$ controls the mass splitting among different KK particles, and $\Lambda R$ defines the number of permitted KK modes below scale $\Lambda$. The perturbativity considerations of the $U(1)_Y$ gauge coupling demands that $\Lambda R \lesssim 40$ in the case of the mUED. Authors of Ref.~\cite{Datta:2012db, Datta:2013xwa} posited that a significantly robust constraint comes from running of the Higgs boson scalar self-coupling ($\lambda$) and also of EW vacuum stability. Throughout this analysis it is assumed that $\Lambda R = 5$.

The LHC is a proton-proton collider at which the first KK level strongly interacting excited states (i.e. excited gluons $g^{1}$ and quarks $Q^{1}/q^1$) will be copiously produced due to color factor and strong coupling. These particles eventually decay and these decays, in the present model, will be governed by two different decay mechanisms: the KK number preserving decay (KKPD) and the Gravity induced decay (GID). In the discussion below, we provide a concise examination of the decay process of the first level KK excitations in the present model.  
\subsection{Decays}
\subsubsection*{Kaluza-Klein Number Preserving Decays (KKPD)}
Preservation of KK-number (and KK-parity) enables the first level KK modes to be produced only in pairs at collider experiments. Tree-level masses of KK excitations, which can be determined by the compactification scale $R$, are almost the same for all KK excitations (except for the KK excitation of top quark and the Higgs boson) of the SM particles. Therefore, at tree-level the mass spectrum is quite degenerate. The mass splitting among different KK particles at level-$n$ can be achieved by inclusion of loop corrections which introduce the cutoff scale $\Lambda$. This enables the decay of first level KK particles to other lighter first level KK particles plus one or more respective SM particle(s) assuming it is kinematically permitted. The decay cascade terminates at the lightest massive KK-particle (LKP) which is $\g^{1}$ in the context of the current model.
\begin{figure}[h!]
\begin{center}
\includegraphics[width=0.9\linewidth, angle=0]{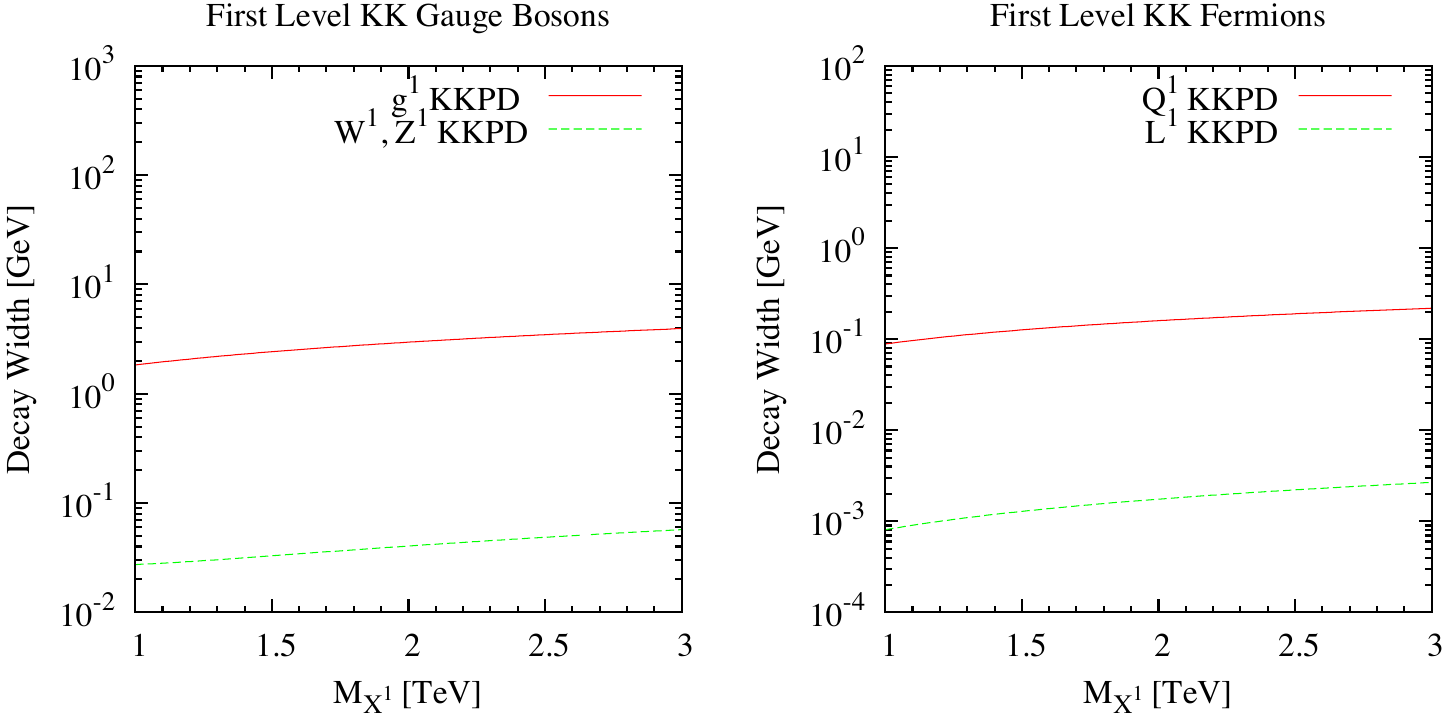}\\
\caption{KK-number preserving (KKPD) decay widths of the first level Kaluza-Klein (KK) gauge bosons (left-hand pane) and fermions (right-hand pane) in terms of decaying particle's mass $M_{X^{1}}$. $M_{D}$ and $\Lambda$ are both set to $5$~TeV. The LKP ($\g^{1}$) does not have any KKPD decay mode, thus only decays gravitationally.}
\label{fig:kkpd} 
\end{center}
\end{figure}
The excited gluon ($g^{1}$) is the most massive among the first level KK particles, succeeded by the KK doublet ($Q^1$) and singlet ($q^1$) quarks. The succeeding states are the electroweak gauge bosons ($W^{\pm 1}/Z^{1}$), lepton doublets ($L^{1}$), and singlets ($e^1$), in that order of decreasing mass. The KK excitation of gluon $g^1$, decays to first level KK quarks (both doublets ($Q^1$) and singlets ($u^1, d^1$)) via nearly equal branching ratios. The singlet KK quarks, having only a single decay channel, decay to the LKP and the corresponding SM quark. The doublet KK quarks decay into first level KK partners of the SM electroweak gauge bosons $W^{\pm 1}$ or $Z^1$ and an associated SM quark with high probability. Since it is kinematically closed, the hadronic decays of the first level electroweak gauge bosons are not realized. They decay into first level KK lepton flavors and the corresponding SM lepton with similar branching ratios. The KK leptons eventually decay into a SM lepton and $\g^1$. Of particular significance is the fact that the KKPD widths heavily rely on the mass gap among the same level KK excitations, thereby exhibiting dependence on the cutoff scale $\Lambda$ and the size of small extra dimension $R$. In contrast, the parameters $\delta$ and $M_{D}$ do not directly contribute to the KKPD. In Fig.~\ref{fig:kkpd}, we provide KKPD widths for the first level KK gauge bosons (left-hand pane), quark, and lepton doublets (right-hand pane) in the mass range of $1-3$ TeV in terms of decaying particle's mass. In producing Fig.~\ref{fig:kkpd} we assumed that the higher dimensional Planck mass $\MD$ and the cutoff scale $\Lambda$ are both fixed to $5$ TeV. Since $\g^1$ is the least massive state in the spectrum, it has no decay mode that preserves KK-number. It only decays gravitationally to photon or $Z$ and KK graviton excitation $\hat{G}^{\vecn}$. The gravity induced decays of the first level KK modes are discussed below.
\subsubsection*{Gravity Induced Decays (GID)}
Distinctive decay mechanism for the KK particles in the present model is realized through interaction between gravity and matter fields. The translational invariance along the coordinate $y$, which provides the preservation of KK-number, is broken down by means of specific positioning of $3$-brane in extra dimensions. Therefore, we note that the KK-number and KK-parity are not intact in the fat-brane UED models because of the Gravity Induced Decays (GID). This allows direct decay of first level KK particles into a gravity excitation $\hat{G}^{\vecn}$ and the corresponding SM particle, hence making the collider phenomenology of the present model quite interesting and different from that of UED models. Besides, by means of gravity induced decays, the direct decay of the LKP into photon or $Z$-boson plus gravity excitations is also realized. Therefore, the LKP which is stable in the mUED, is no longer stable in the fat-brane realizations of UED models. The GID widths of excited KK particles into gravitons and the SM particle were formerly determined. We refer interested readers to Ref.~\cite{Macesanu:2003jx, Macesanu:2005jx, Ghosh:2012zc}.
\begin{figure}[h!]
\begin{center}
\includegraphics[width=0.9\linewidth, angle=0]{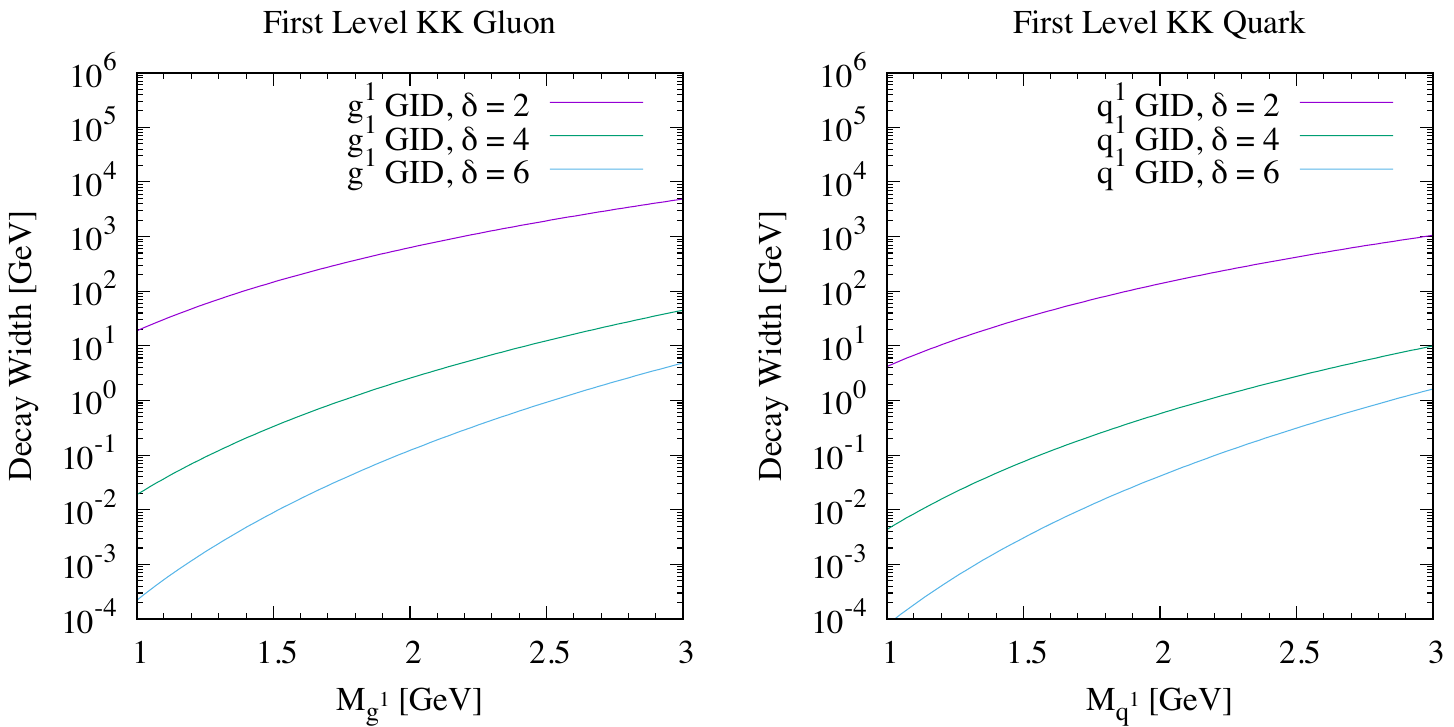}\\
\caption{Gravity induced decay (GID) width of the first level Kaluza-Klein (KK) gluons (left-hand pane) and quarks (right-hand pane) in terms of decaying particle's mass for three different values of large extra dimensions ($\delta=2, 4, 6$). $M_{D}$ and $\Lambda$ are both set to $5$~TeV.}
\label{fig:gid} 
\end{center}
\end{figure}
\begin{figure}[h!]
\begin{center}
\includegraphics[width=0.9\linewidth, angle=0]{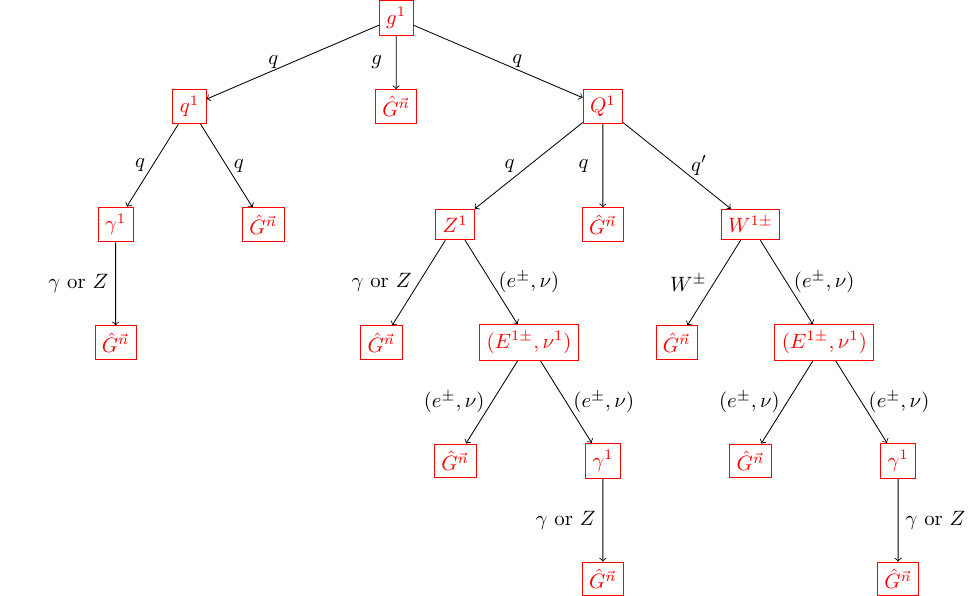}
\caption{The dominant decay pathway of the first level KK gluon ($g^{1}$) and quarks (doublet $Q^{1}$ and singlet $q^{1}$) for $\delta=2,4$. $\hat{G}^{\vecn}$ is the gravity excitation (graviton $h^{\vecn}$, gravi-vector $A^{\vecn}$, and gravi-scalar $\phi^{\vecn}$).}
\label{fig:decay_chain_D24}   
\end{center}
\end{figure}
\begin{figure}[h!]
\begin{center}
\includegraphics[width=0.5\linewidth, angle=0]{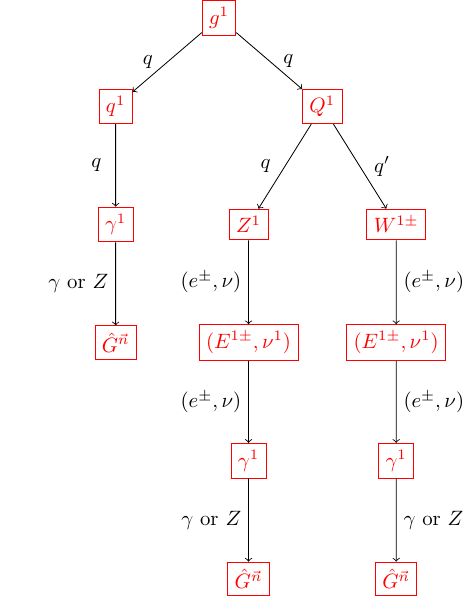}
\caption{The dominant decay pathway of first level KK gluon ($g^{1}$) and quarks (doublet $Q^{1}$ and singlet $q^{1}$) for $\delta=6$. $\hat{G}^{\vecn}$ is the gravity excitation (graviton $h^{\vecn}$, gravi-vector $A^{\vecn}$, and gravi-scalar $\phi^{\vecn}$).}
\label{fig:decay_chain_D6} 
\end{center} 
\end{figure}
We provide the widths of GID in Fig.~\ref{fig:gid} for the first level KK gluons (left-hand pane) as well as for quarks (right-hand pane) for a varying mass of decaying particle in $M_X=1-3$ TeV range. In producing Fig.~\ref{fig:gid} we fixed $\MD=5$ TeV and $\Lambda=5$ TeV. As argued above the KKPD width is insensitive to $\delta$, the number of large extra dimensions. The GID, contrarily, has high dependence to this parameter. The GID width of an individual decaying first level KK particle is the maximum for $\delta=2$ and the minimum for $\delta=6$, respectively. This behavior can be attributed to the fact that the mass splitting between different level-$\vecn$ gravity excitations $\hat{G}^{\vecn}$ is directly proportional to $\delta$ for fixed $\Rinv$. For smaller values of $\delta$, one obtains smaller mass splittings in gravity excitations and thus larger number density of KK-gravitons in a given mass range. This ensures a larger(smaller) value for the GID in the case of $\delta=2(6)$. In the case of $\delta=4$ the KKPD and GID widths exhibit comparability. Based on this discussion, in Fig. \ref{fig:decay_chain_D24} and Fig. \ref{fig:decay_chain_D6} we schematically present dominant decay pathways of the first level KK particles for $\delta=2,~4$ and $\delta=6$, respectively. In the following we discuss the effects of these decay mechanisms and possible final states at the LHC experiment. 
\subsection{Collider signals of fat-mUED at the LHC}
Having analyzed the decay pathways of the first level KK excitations, we now proceed to examine the collider phenomenology of the present model within the framework of the LHC experiment. The first level KK quarks and gluons, due to their color factors and strong coupling, are expected to be produced at substantial rates\footnote{The QCD production of the first level KK particles is dominant compared to the electroweak production. Therefore, we neglected the latter production mechanism in the present study.}. The pair-production of these is ensured by the fact that the KK-parity is preserved in the interaction vertices of the KK particles with the SM particles. The production rate is mainly determined through $R$ which fixes the masses of these excitations, though there is a mild dependence on the cut-off scale $\Lambda$ which essentially controls the mass gap among the first level KK excitations. Throughout the analysis we take $\Lambda R = 5$. After the production, the first level KK quarks and gluons decay and the final state topology of resultant signal is dictated by the relative strength of two different decay mechanisms discussed above. Therefore, the question of which final state would be most probable are closely tied to the value of $\delta$. To illustrate this, consider the scenario of $\delta=2$ for which the GID is order of magnitude larger than the KKPD width for a given decaying particle. Provided that the dominant decay pathway for first level KK gluons and quarks (both doublet and singlet) are as given in Fig.~\ref{fig:decay_chain_D24}, $g^{1}$ undergoes a direct decay to SM gluon and level-$\vecn$ KK graviton excitation $\hat{G}^{\vecn}$. In a similar fashion, the first level excited quarks decay to a SM quark plus $\hat{G}^{\vecn}$. Therefore, it is anticipated, in the case of $\delta=2$, that the decay of the first level KK quark and/or gluon pairs produce a final state which is characterized by two hard jets and substantial missing $E_{T}$ at the LHC experiment. The missing transverse energy accompanying jets are due to presence of KK gravity excitations which are too weak to be observed in the detector. The relative strength between two decay mechanisms radically changes and modifies the expected final states at the LHC in the case of $\delta=6$. For this case, one sees a suppression of gravity induced decays over the KKPD. This means that the pair-produced first level KK particles (gluons and quarks) cascade decay to other lighter first level KK particles plus low-$p_{T}$ SM particles (jets, leptons etc.) through KK-number preserving decays. The decay cascade ends with the production of the LKP $\g^{1}$. The LKP has no KK-number preserving decay, hence only decays gravitationally to $\g$ or $Z$ in association with $\hat{G}^{\vecn}$. For that reason, for $\delta=6$ wherein the GID contributes only in the last part of the decay chain, the decay of pair-produced KK excitations of quarks and gluons result in $\g\g$, $\g Z$ or $ZZ + X + \slashed{E}_{T}$ signal where $\slashed{E}_{T}$ is due to the undetected gravity excitations $\hat{G}^{\vecn}$ and $X$ represents the SM particles (jets, leptons etc.) radiated in the cascade. At this point it is of considerable import to state that the decay of $\g^{1}$ occurs within the boundaries of the detector for the set of parameters under discussion herein. We have ascertained that for the mass range of $1$ to $3$ TeV, the GID width for $\g^{1}$ ranges from $\sim 10$ keV to $1$ GeV, thereby facilitating prompt decay of the LKP within the detector. In the intermediate case of $\delta=4$ where the GID and the KKPD are of the similar order of magnitude, as seen in Fig.~\ref{fig:kkpd} and Fig.~\ref{fig:gid}, both decay modes may contribute the decay of excited KK particles. For instance, one parton may cascade decay to lighter first level KK particles, radiating low-$p_{T}$ jets and/or leptons up to the LKP which then further decays gravitationally to $\g$ or $Z$-boson plus $\hat{G}^{\vecn}$. On the other hand, the other parton may undergo a direct decay to an associated SM particle and $\hat{G}^{\vecn}$. Therefore, the decay of pair-produced gluons/quarks can produce jets$+ \g + \slashed{E}_{T}$ final state wherein $\slashed{E}_{T}$ is due to undetected gravity excitations.     
 
After discussing the decay mechanisms and possible phenomenological final state signals of the fat-mUED model at the collider experiments, we now move to present implications of two recent searches by the ATLAS experiment on the parameters of fat-mUED model. In this work, we confronted the fat-mUED model with the ATLAS searches on jets$+ \g$ and multi-jet plus $\slashed{E}_{T}$ final states at $13$ TeV LHC with $136$ fb$^{-1}$ integrated luminosity. In the subsequent sections below, we present a succinct overview of these aforementioned ATLAS searches.
\subsection{Multi-jet $ + \slashed{E}_{T}$ search by the ATLAS Collaboration}\label{sec:multijet}
\begin{table}[t!]
  \centering
  \begin{tabular}{|c||c|}
    \hline\hline 
\textbf{Lepton Veto} & $N_{e/\mu} = 0$ with $p_{T}^{e/ \mu}>7/6$ GeV \\
\hline
$\slashed{E}_{T}$ & $>300$ GeV\\
\hline
$p_T(j_1)$ & $> 200$ GeV\\
\hline 
$p_T(j_2)$ & $> 50$ GeV\\
\hline
$\Delta \phi (j_{1,2,(3)},\vec{p}_{T}^{~miss.})_{\text{min.}}$ & $> 0.2(0.4)$ rad.\\
\hline
$m_{\text{eff.}}$ & $>0.8$ TeV\\
\hline
\hline
  \end{tabular}\caption{List of preselection criteria applied in ATLAS multi-jet plus missing transverse energy search \cite{ATLAS:2020syg}. $\Delta \phi (j_{1,2,(3)},\vec{p}_{T}^{~miss.})_{\text{min.}}$ is the minimum azimuthal separation in radians between the missing energy vector $\vec{p}_{T}^{~miss.}$ and $p_{T}$ of up to three hardest jets. $m_{\text{eff.}}$ is calculated by $m_{\text{eff.}} = \slashed{E}_{T} + \sum_{i=all}p_{T}(j_{i}>50\text{~GeV})$. The minimum value on $\Delta \phi (j_{1,2,(3)},\vec{p}_{T}^{~miss.})_{\text{min.}}$ in parenthesis corresponds to cut used in Ref.~\cite{ATLAS:2020syg}. For validation purposes (see Section \ref{sec:validation}) we used the cut-flow Table (Table 17 in Appx. B) provided in Ref.~\cite{ATLAS:2019vcq}.}
\label{tbl:preselection}
\end{table}

The ATLAS Collaboration at the LHC has recently shared the findings \cite{ATLAS:2020syg} of a detailed search on multi-jet ($2-6$ jets) plus missing transverse energy final state from $pp$ collisions at $\sqrt{s} = 13$ TeV LHC with $139$ fb$^{-1}$ integrated luminosity of data. The aim of the ATLAS search was to find the traces of strongly produced sparticles (squarks and gluinos) in R-parity conserving SUSY models at the LHC. The ATLAS Collaboration, due to the absence of excess number of events over the SM background, has placed model-independent $95\%$ C.L. upper limits on NP contribution to visible cross-sections ($\sigma_{\text{vis.}}$)\footnote{The visible cross-section ($\sigma_{\text{vis.}}$) is the product of production cross-section ($\sigma_{\text{prod.}}$), acceptance $\mathcal{A}$, and efficiency $\epsilon$, by definition.} in various signal regions (SRs). Although the limits obtained in ATLAS search were interpreted in constraining SUSY models, one may use these null results to constrain any BSM model which predicts a similar final state, \textit{viz., events with high jet-multiplicity  and missing energy}, at the LHC. In the present study, an analogous exercise is performed for the fat-mUED model. We translated the above-mentioned upper limits on the visible cross-sections in SRs into limits on the parameters of the fat-mUED model. In the following we elaborate on the ATLAS multi-jet search. 

In ATLAS multi-jet$+\slashed{E}_{T}$ search, anti-kT jet clustering algorithm \cite{Cacciari:2008gp} via jet radius parameter $\Delta R = 0.4$ is used. Only reconstructed jets with $p_{T}^{j} > 20$ GeV and $|\eta^{j}|<2.8$ are passed to subsequent stages of the analysis. Lepton candidates (both electrons and muons) are demanded to satisfy $p_{T}^{e(\mu)} > 7(6)$ GeV and to be within $|\eta^{e(\mu)}|<2.47(2.7)$ rapidity window. Following jet/lepton identifications, any electron/muon candidate present within a distance $\Delta R = \text{min}(0.4, 0.04 + 10 \text{~GeV}/p_{T}^{e/\mu})$ of any jet candidate is discarded. The determination of missing transverse momentum vector $\vec{p}_{T}^{~miss.}$ and its magnitude $\slashed{E}_{T}$ rely on all reconstructed jets, leptons, and calorimeter clusters that are not associated with any of these aforementioned objects. Following the reconstruction of different objects, a set of preselection criteria, listed in Table \ref{tbl:preselection}, is applied to events:
 
Events containing an isolated electron (muon) with $p_T>7(6)$ GeV are vetoed. Only events with a leading jet\footnote{Jets are ordered in momentum with $p_{T}^{j_1}$ being the highest, accordingly.} with $p_T^{j_1}>200$ GeV, and a sub-leading jet with $p_T^{j_2}>50$ GeV are passed to further analysis. $m_{eff}$, a powerful discriminator in high mass scale searches, is calculated as the scalar sum of $\slashed{E}_{T}$ and the transverse momenta of all jets with $p_{T}>50$ GeV. $m_{eff}>800$ GeV and a sufficiently high missing transverse energy $\slashed{E}_{T} > 300$ GeV requirements are also applied to events passing preselection criteria. Additionally, events are required to satisfy the condition on azimuthal separation between missing momentum vector $\vec{p}_{T}^{~miss.}$ and the momenta of up to three hardest jets, $\Delta \phi(j_{1,2,(3)},\vec{p}_{T}^{~miss.})_{\text{min.}}>0.4$.\\
In the study of Ref.\cite{ATLAS:2020syg}, the ATLAS Collaboration defined ten signal regions (SRs) characterized by increasing number ($2$ to $6$) of jets and also with minimum values on $m_{eff}$ variable. The ATLAS Collaboration reported an absence of a statistically significant number of events above the Standard Model background, leading to the establishment of model-independent 95$\%$ C.L. upper limits on the visible cross-section contribution $\sigma_{\text{vis.}}$ arising from any BSM scenario across all signal regions (SRs).

We have used these model independent limits to restrict the parameters of the fat-mUED model. In Table \ref{tbl:multijet_cuts}, we list cuts applied by the ATLAS Collaboration in the analysis and upper limits on visible cross-sections for di-jet signal regions\footnote{We omitted SRs with higher jet multiplicities as they do not offer higher exclusion limits than those found in SRs with two jets.}.
\begin{table}[t!]
  \centering
  \begin{tabular}{|c||c|c|c|}
    \hline\hline
    & \multicolumn{3}{c|}{\textbf{Signal Region}}\\
\hline\hline
\textbf{Cuts} & \textbf{2J-1600}&\textbf{2J-2200}&\textbf{2J-2800}\\
\hline\hline
$N_{j}$ &\multicolumn{3}{c|}{$\geq 2$} \\
\hline
$p_{T}(j_{1})$ [GeV] & $> 250$&$> 600$&$> 250$\\
\hline
$p_{T}(j_{i=2,3,...,N_{j_{min}}})$ [GeV] & $> 250$  & $> 50$ & $> 250$\\
\hline
$|\eta(j_{i=1,2,...,N_{j_{min}}}|$ &$<2.0$& $<2.8$ & $<1.2$\\
\hline
$\Delta \phi(j_{1,2,(3)},p_{T}^{miss.})_{min}$  & $>0.8$&$>0.4$&$>0.8$\\
\hline
$\Delta \phi(j_{i>3},p_{T}^{miss.})_{min}$  & $>0.4$&$>0.2$&$>0.4$\\
\hline
$\slashed{E}_{T} /\sqrt{H_{T}}$ [GeV$^{-1/2}$] &\multicolumn{3}{c|}{$>16$} \\
\hline
$m_{\text{eff.}}$ [TeV]& $>1.6$& $>2.2$&$>2.8$\\
\hline
\hline
$\sigma_{\text{vis.}}$ [fb]&\bf 1.47 & \bf 0.78 & \bf 0.14\\
\hline
\hline
  \end{tabular}\caption{Kinematic cuts and signal region definitions of the ATLAS multi-jet search \cite{ATLAS:2020syg} and $95\%$ C.L. model-independent upper limits on visible cross-sections $\sigma_{\text{vis.}}$ for \textbf{2J-1600}, \textbf{2J-2200}, and \textbf{2J-2800} signal regions. $\Delta \phi (\text{jets},~ \vec{p}_{T}^{miss.})_{min}$ represents the smallest azimuthal separation between the missing momentum vector $\vec{p}_{T}^{miss.}$ and the momenta of the hardest jet(s). $H_{T}$ is defined as $H_{T} = \sum_{i=all} p_{T}(j_{i}> 50\text{~GeV})$. In a similar way, $m_{\text{eff.}}$ is calculated as $m_{\text{eff.}} = \slashed{E}_{T} +\sum_{i=all} p_{T}(j_{i}> 50\text{~GeV})$.}
\label{tbl:multijet_cuts}
\end{table}
\subsection{$\text{Jets} + \g$ plus missing energy search by the ATLAS Collaboration}\label{sec:photon}
Final states comprising photons, jets and a large missing transverse energy signature can also be realized in the case of gauge-mediated SUSY breaking (GMSB) scenarios.  In GMSB, the pair-produced strongly interacting particles (i.e. gluinos and squarks) cascade decay into other lighter SUSY particles, radiating jets. The decay cascade proceeds until the next-to-lightest supersymmetric (NLSP) particle, the lightest neutralino which further decays into an ultra-light gravitino and a $\g /Z /h$. The gravitino is the lightest massive supersymmetric particle (LSP), so it escapes from detection and introduces a missing transverse energy in events. Therefore, in the GMSB scenario the decay of pair-produced gluinos and/or squarks may lead to a final state topology of high-$p_{T}$ photon(s) with jets and large missing transverse energy carried off by the gravitino at the LHC. For the case of current scenario, the gravity induced decay of ($\g^{1}$) into $\g$ and $\hat{G}^{\vecn}$ produces a high-$p_{T}$ photons as well. Therefore, along with the jets radiated from KKPD of the first level KK gluons and/or quarks, one may produce a similar final state \textit{viz.,} $\text{jets}+\g$ and $\slashed{E}_{T}$ in fat-brane UED model. This final state topology would be more pronounced especially when the GID participates in the final step of the decay chain, generating events with photon presence, corresponding to $\delta=4,6$ cases.     

The ATLAS Collaboration \cite{ATLAS:2022ckd} searched for a sign of GMSB scenario in  $\text{jets} + \g + \slashed{E}_{T}$ signal in $pp$ collisions at $13$ TeV LHC with $139$ fb$^{-1}$ integrated luminosity. In the absence of excess number of events over the SM prediction, the ATLAS Collaboration placed upper limits on visible cross-sections $(\sigma_{\text{vis.}})$ at $\% 95$ C.L. on any BSM contribution to $\text{jets} + \g + \slashed{E}_{T}$ final state for different SRs. We translated these limits on the visible cross-sections to constrain the parameters of the fat-mUED model. The search strategy employed by the ATLAS is discussed in the following.

The ATLAS Collaboration defined three SRs, named as SRL, SRM, and SRT, which are mainly characterized by increasing jet multiplicity in an event. Algorithm employed in reconstruction of objects such as jet, leptons and missing transverse energy $\slashed{E}_{T}$ are similar to those of multi-jet analysis discussed in Section \ref{sec:multijet}. SR jets are required to have $p_{T}>30$ GeV and be within rapidity range of $|\eta|<2.5$. An extra requirement of $p_{T}>50$ GeV for SR photons are employed. Due to the considerable mass scale of SUSY particles, \textit{viz.,} gluinos and squarks, being explored in ATLAS search, an expectation of high momentum for visible particles is inevitable. $H_{T}$ is calculated as $H_{T} = p_{T}^{leading}(\g)+\sum_{i=all}p_{T}(j_i)$ and events are vetoed if a cut on $H_{T}$ is not satisfied. Moreover, $R_{T}^{4}$, defined as the ratio of the scalar sum of the transverse momenta of the four hardest jets and the scalar sum of transverse momenta of all jets in the event, is useful in discriminating the SM background from the signal in events with a high number of jets. Hence, it is used in SRL and SRM signal regions, both of which contain a high number of jets compared to SRT. We list event selection criteria and the $95\%$ C.L. upper limits on $\sigma_{\text{vis.}}$ in SRs for $\text{jets} + \g + \slashed{E}_{T}$ signal of $13$ TeV ATLAS search \cite{ATLAS:2022ckd} in Table \ref{tbl:photon_cuts}.          
\begin{table}[t!]
  \centering
  \begin{tabular}{|c||c|c|c|}
    \hline\hline
    & \multicolumn{3}{c|}{\textbf{Signal Region}}\\
\hline\hline
\textbf{Requirements} & \textbf{SRL}&\textbf{SRM}&\textbf{SRT}\\
\hline\hline
$N_{photons}$ &\multicolumn{3}{c|}{$\geq 1$} \\
\hline
$p_T^{\text{leading}}(\g)$ [GeV]  & $> 145$&$> 300$&$> 400$\\
\hline
$N_{leptons}$ & \multicolumn{3}{c|}{$0$}\\
\hline
$N_{jets}$ &\multicolumn{2}{c|}{$\geq 5$} & $\geq 3$\\
\hline
$\Delta \phi(jet,E_{T}^{miss.})$ [rad.]  & \multicolumn{3}{c|}{$>0.4$}\\
\hline
$\Delta \phi(photon,E_{T}^{miss.})$  [rad.]  & \multicolumn{3}{c|}{$>0.4$}\\
\hline
$E_{T}^{miss.}$ [GeV]  & $>250$& $>300$ & $>600$ \\
\hline
$H_{T}$ [GeV]& $>2000$ &\multicolumn{2}{c|}{$>1600$}\\
\hline
$R_{T}$ &\multicolumn{2}{c|}{$<0.90$}& - \\
\hline
\hline
$\sigma_{\text{vis.}}$ [fb] &\bf 0.034 & \bf 0.022 & \bf 0.054\\
\hline
\hline
\end{tabular}
\caption{Kinematic cuts and signal region definitions of the ATLAS \cite{ATLAS:2022ckd} $\text{jets}+\g$ plus $\slashed{E}_{T}$ search and $95\%$ C.L. model-independent upper limits on visible cross-sections ($\sigma_{\text{vis.}}$) for \textbf{SRL}, \textbf{SRM}, and \textbf{SRT} signal regions. $H_{T}$ is given as $H_{T} = p_{T}^{leading}(\g)+\sum_{i=all}p_{T}(j_i)$. $R_{T}^{4}$ is the ratio of the scalar sum of the transverse momenta of the four hardest jets and the scalar sum of transverse momenta of all jets in the event. Additionally, $\Delta \phi(X,E_{T}^{miss.})$ where $X=$ jet/photon represents the azimuthal separation between $X$ and $\vec{\slashed{E}_{T}}$.}  
\label{tbl:photon_cuts} 
\end{table}

\subsection{Event simulation, object reconstruction and validation} 
\label{sec:validation}
In this section, we discuss the specifics of the signal event simulation, object reconstruction and the validation of our analysis. \textbf{PYTHIA} \cite{Sjostrand:2006za} with the implementation of mUED \cite{ElKacimi:2009zj} is used in generating parton-level events of the first level KK quarks and gluons produced in pairs at $13$ TeV. We used the \textbf{NNPDF23LO} \cite{NNPDF:2014otw} parton distribution functions with keeping factorization ($\mu_F$) and renormalization ($\mu_R$) scales fixed at $\sqrt{\hat{s}}$. In the `fat-brane' implementation of mUED, GID widths are assumed to be smaller than KKPD widths which may be true in the case of $\delta=6$ but not in the case of $\delta=2,~4$ as can be seen comparing Fig.~\ref{fig:kkpd} and Fig.~\ref{fig:gid}. The aforementioned implementation of mUED, also, includes GID of only the LKP into photon and gravity excitations ($\g^{1} \to \g + \hat{G}^{\vecn}$) and not that of other heavier first level KK excitations. Additionally, contrary to the SM, the mixing angle between $B_{\mu}^{1}$ and $W_{\mu}^{3}$ is minuscular as discussed previously, so $\g^{1}$ is basically the first level KK excitation of $B_{\mu}$. The PYTHIA implementation of mUED does not consider $\g^{1} \to Z + \hat{G}^{\vecn}$ decay and assumes Br$(\g^{1} \to \g + \hat{G}^{\vecn})=100\%$. We accommodated both GID channels of the LKP and GID of other heavier first level KK excitations to our analysis by modifying PYWIDTH subroutine in \textbf{PYTHIA} accordingly. The details of ATLAS searches given in \cite{ATLAS:2020syg} for multi-jet plus $\slashed{E}_{T}$ and for $\text{jets} + \g$ plus $\slashed{E}_{T}$ \cite{ATLAS:2022ckd} final states are firmly applied in the event simulation and in the construction of physical objects (\textit{viz., jets, leptons, photons, {\normalfont and} $\slashed{E}_{T}$}) in our analysis. The jets are reconstructed with anti-kT jet clustering algorithm in \textbf{FastJet} \cite{Cacciari:2011ma}. It is imperative to verify the congruence of our analysis with the one executed by the ATLAS at this juncture. In the case of multi-jet search the ATLAS Collaboration presents the cut-flow Table (Table $17$ in Ref.\cite{ATLAS:2019vcq}) for gluino pair-production followed with direct decay into quark-antiquark pair plus a neutralino ($\tilde{g} \to q\bar{q}\tilde{\chi}^{1}_{0}$) for $m_{\tilde{g}}=2.2$ TeV and $m_{\tilde{\chi}^{1}_{0}} = 0.6$ TeV. The analysis technique of the ATLAS in Ref.~\cite{ATLAS:2019vcq} is subsequently used in Ref.~\cite{ATLAS:2020syg} with minor modifications. The formerly determined bounds on SUSY particles for simplified models are then updated accordingly. We used Ref.~\cite{ATLAS:2019vcq} for validating our analysis; however, the upper bounds on visible cross-sections in the signal regions presented in Ref.~\cite{ATLAS:2020syg} were used to obtain the exclusion limits on the fat-mUED model.
For validation purpose, we generated the gluino-gluino production followed by their direct decay to quark pairs and a neutralino with above-mentioned masses in \textbf{PYTHIA}. The initial state radiations (ISR), decays and showering are also performed with \textbf{PYTHIA}. In Table \ref{tbl:comparison} we present the cut-efficiencies supplied by the ATLAS \cite{ATLAS:2019vcq} (2nd column) and the results of our simulation (3rd column) for comparison purpose.
\begin{table}[t!]
  \centering
  \begin{tabular}{|c||c|c|}
    \hline\hline
\multirow{2}{*}{\textbf{Process}} & \multicolumn{2}{c|}{$pp \rightarrow \tilde{g}\tilde{g}$ with direct decay $\tilde{g} \rightarrow q\bar{q}\tilde{\chi}_{1}^{0}$} \\
& \multicolumn{2}{c|}{$m_{\tilde{g}} = 2.2$ TeV and $m_{\tilde{\chi}_{1}^{0}} = 0.6$ TeV}\\
\hline
\multirow{2}{*}{\textbf{Cuts}} & \multicolumn{2}{c|}{Absolute efficiency in $\%$} \\\cline{2-3}
& ATLAS (Table 17 in Appx. B of \cite{ATLAS:2019vcq}) & Our Simulation \\ \hline \hline
preselection$+N_j \geq 2$ & 100.0 & 99.9 \\
$N_j \geq 4$ & 92.9 & 96.6 \\
$\Delta \phi(j_{1,2,(3)},\vec{\slashed{p}}_{T})_{min}>0.4$ rad.&  77.6 & 80.6\\
$\Delta \phi(j_{i>3}, \vec{\slashed{p}}_{T})_{min}>0.2$ rad. &  69.1 & 71.9\\
$p_{T}(j_4) > 100$ GeV & 61.3 &56.5\\
$|\eta(j_{i=1,2,3,4})|<2.0$ &  55.7 & 51.7\\
Aplanarity$>0.04$ & 38.7 & 40.1 \\
$\slashed{E}_{T} / \sqrt{H_{T}} >16$ GeV$^{1/2}$ & 24.1 & 25.2\\
$m_{eff}>1.0$ TeV & 24.1 &25.2\\
\hline\hline
  \end{tabular}
\caption{Cut-flow table of \textbf{4J-1000} signal region as presented in Ref.~\cite{ATLAS:2019vcq} of ATLAS multi-jet search (2nd column) along with our numbers for the same signal region (3rd column). The aplanarity is given by $A=\frac{3}{2} \lambda_{3}$, in which $\lambda_3$ represents the minimum eigenvalue of the normalized momentum tensor of jets in the event.}  
\label{tbl:comparison}
\end{table}    
The results shown in Table \ref{tbl:comparison} show that the analysis strategy we followed is in good agreement with the ATLAS analysis. Below we present the exclusions obtained by performing such an analysis.     

\section{Exclusion limits on fat-mUED model parameters}\label{sec:Exclusions}
\begin{figure}[!htb]
\includegraphics[width=0.5\linewidth]{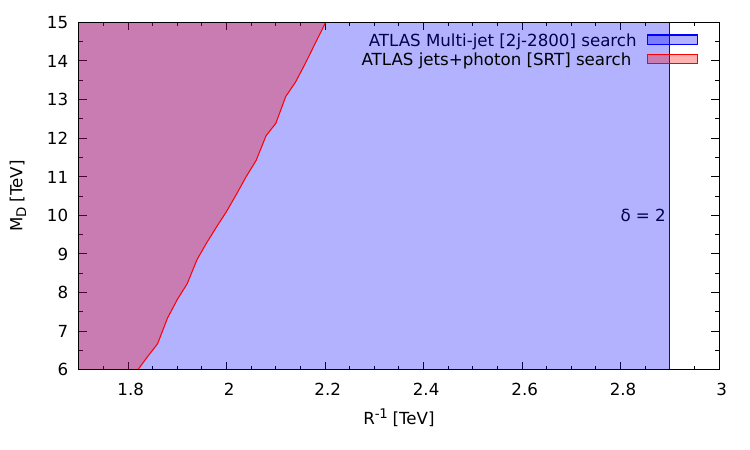}
\includegraphics[width=0.5\linewidth]{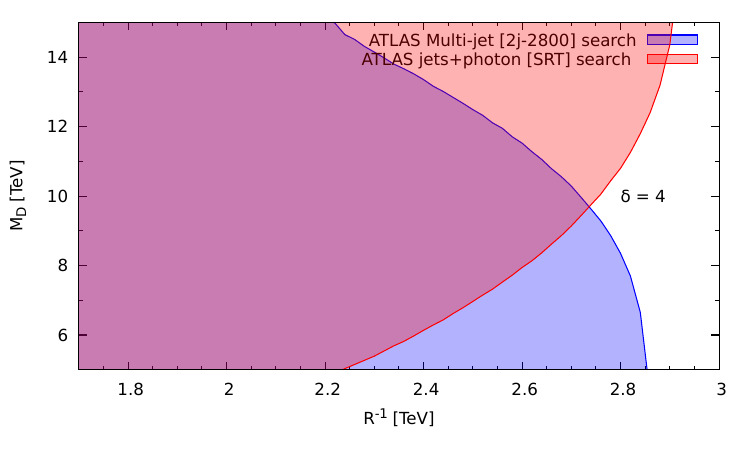}
\begin{center}
 \includegraphics[width=0.5\textwidth]{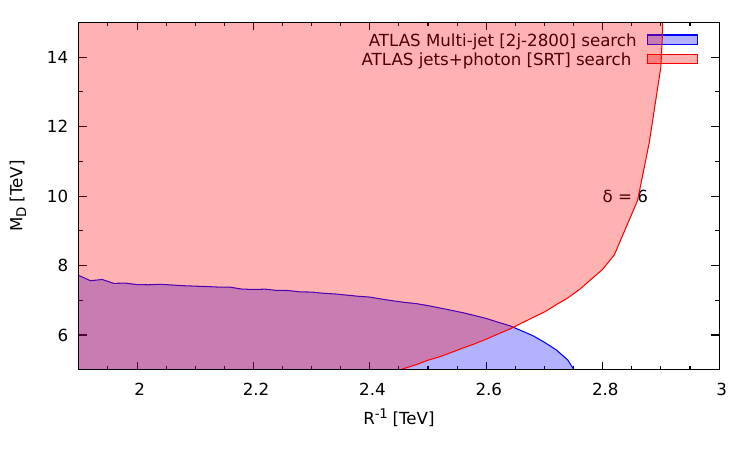}\\
\end{center}
\caption{Exclusion regions of fat-mUED model at $95\%$ C.L. for $\delta=2$ (top left-hand pane), $\delta=4$ (top right-hand pane), and $\delta=6$ (bottom pane). Blue (red)-shaded area in $\Rinv-\MD$ planes corresponds to exclusions from the ATLAS multi-jet (jets +$\g$) plus missing transverse energy search, correspondingly. We set $\Lambda \Rinv = 5$ throughout exclusion plots.   
}\label{fig:exclusion}
\end{figure}

In this section, we present the exclusion limits at $95\%$ C.L. on the fat-mUED model coming from ATLAS multi-jet and $\text{jets} + \g + \slashed{E}_{T}$ searches at $13$ TeV LHC. The collider phenomenology of the fat-mUED model is fully determined by the fundamental parameters of the model, specifically the compactification scale $\Rinv$, the number of large extra dimensions $\delta$, and the $(4+\delta)$D Planck mass $\MD$. There is also a mild dependence on the cut-off scale $\Lambda$ which we set to $5$ TeV throughout the analysis. Since $\Rinv$ and $\Lambda$ determine the scale at which the masses of first level  KK excitations reside, the pair-productions of KK particles mainly depend on these parameters. On the other hand, $\delta$ and $\MD$ control the mass differences among KK gravitons, thereby affecting the density of KK states and the decay widths induced by gravity. Henceforth, the final state signal topology strongly depends on these parameters. In order to obtain exclusions from aforementioned ATLAS searches on the fat-mUED model we scanned the fundamental parameters of the model in $\Rinv=[1-3]$ TeV and $M_{D} = [5-15]$ TeV range with $20$ GeV and $100$ GeV step sizes, respectively. The scan is repeated for $\delta=2,4,6$. The obtained fat-mUED cross-section contributions to each defined SR in ATLAS searches are then compared with $95\%$ C.L. limits reported in Ref.\cite{ATLAS:2020syg} and in Ref.\cite{ATLAS:2022ckd}. We present the exclusion results in $\Rinv-M_{D}$ plane for $\delta=2$ (top left-hand pane), $\delta=4$ (top right-hand pane), and $\delta=6$ (bottom pane) in Fig.~\ref{fig:exclusion}. Below we present our limits for $\delta=2,4$ and $\delta=6$ in detail:\\

\begin{enumerate}  
  
\item \textbf{Bounds on ($\Rinv-M_{D}$) plane for $\delta=2$:} As we discussed above, the value of $\delta$ mainly specify the mass splitting among KK gravitons and hence the density of these states. For that reason, the gravity induced decay (GID) width is under direct control of $\delta$. For $\delta=2$, as can be seen comparing Fig.~\ref{fig:kkpd} and in Fig.~\ref{fig:gid}, in the mass range of the decaying particle, specifically at $M_{X^{1}}=[1-3]$ TeV, the width of the decay induced by gravity surpasses that of the KK-number preserving width. As a consequence, for $\delta=2$ case, the pair-produced first level KK excitations (the first level quarks and gluons) directly decay to an associated SM particle (quark or gluon), radiating gravity excitation with high probability. Consistent with the decay characteristics we found that di-jet plus missing transverse energy signal is more pronounced in excluding the fat-mUED model. We observed the highest exclusion in $2$j$-2800$ SR among all other di-jet signal regions presented in ATLAS multi-jet search \cite{ATLAS:2020syg} and the exclusion is independent of $\MD$ in the range we scanned in this work. In particular, we found that $\Rinv < 2900$ GeV is excluded from ATLAS di-jet $+ \slashed{E}_{T}$ signal in $M_{D}=[5,20]$ TeV range as shown in Fig.~\ref{fig:exclusion} (top left-hand pane). Alternatively, in the case of $\text{jets} + \g$ plus $\slashed{E}_{T}$ search, we found that only a minuscule part of the parameter domain of the model is excluded compared to multi-jet search. We observed that the exclusion is more pronounced in high-$\MD$ and low-$\Rinv$ area. This feature can be understood by the fact that the KKPD width is small compared to GID width in low-$\MD$ and high-$\Rinv$ region. Therefore, smaller number of photons from a cascade of KK gluons and/or quarks contribute to photonic final state. In particular, $\Rinv < 1850 (2200)$ GeV for $\MD < 6 (15)$ TeV are solely excluded by the ATLAS photonic signature search \cite{ATLAS:2022ckd} at $\%95$ C.L. alone. \\
  
\item \textbf{Bounds on ($\Rinv-M_{D}$) plane for $\delta=4$:} This scenario constitutes a situation in which the GID and the KKPD widths attain a state of comparability, and therefore the interplay between different decay mechanisms becomes more evident on collider phenomenology of the fat-mUED model. In low-$\MD$ and high $\Rinv$ region the GID is the main mechanism for decay of pair-produced first level KK gluons and quarks than that of KKPD. In contrast, within the region of parameters that is characterized by high-$\MD$ and low-$\Rinv$ the KKPD prevails over the GID, and the first level KK quarks and/or gluons produced in pairs undergo a cascade decay to the LKP along with the corresponding SM particles. Finally, the LKP decays to $\g$ or $Z$-boson plus gravity excitation. The consequence of this interplay between two decay mechanisms can be clearly seen in Fig.~\ref{fig:exclusion} (top right-hand pane). In particular, we found that the ATLAS di-jet plus missing transverse energy search excludes $\Rinv < 2850(2200)$ GeV for $M_{D}=5(15)$ TeV at $\%95$ C.L. Similarly, $\Rinv < 2250(2950)$ GeV is excluded for $M_{D}=5(15)$ TeV at $\%95$ C.L. solely by the ATLAS $\text{jets} + \g$ plus missing transverse energy search. \\ 

\item \textbf{Bounds on ($\Rinv-M_{D}$) plane for $\delta=6$:} In the case of $\delta=6$, the KKPD is the main decay mechanism of the first level KK excitations than the GID. As a result of this, the cascade decay of pair-produced first level KK modes of quarks and gluons results in the final states encompassing at least one photon. We found the similar behavior, as presented in Fig.~\ref{fig:exclusion}, in the exclusion regions of the parameters of the fat-mUED model. In particular, we found that $\Rinv <2450(2950)$ GeV is excluded for $\MD=5(15)$ TeV at $\%95$ C.L. by the ATLAS jets $+$ photon and missing transverse energy search. Although the decay width of first level KK excitations is overwhelmed by the KKPD, there is a small part of the parameter region of the model to which ATLAS di-jet search is also sensitive to. In this low-$\MD$ and relatively high-$\Rinv$ region, the first level KK quarks and gluons undergo direct decay to the associated SM particle accompanied by KK excitations of graviton. We found that ATLAS di-jet search excluded at $\MD=5$ TeV for $\Rinv <2750$ GeV at $\% 95$ C.L and the di-jet search is only effective for $\MD<7.5$ TeV at most.\\
  
\end{enumerate}       

\section{Summary and Conclusions}\label{sec:Conclusion}
In the present study, we analyzed the particle collider phenomenology of the fat-mUED model in conjunction with the recent search results from the ATLAS Collaboration. Previously the LHC phenomenology of the model is studied by Ref.~\cite{Ghosh:2012zc} focusing on $\g\g$ signal with $3.1$ pb$^{-1}$ of data results at $7$ TeV energy, and recently in Ref.~\cite{Ghosh:2018mck} with ATLAS $\g\g$ and also multi-jet searches with $36.1$ fb$^{-1}$ at $13$ TeV energy. The ATLAS Collaboration lately communicated the findings of multi-jet plus $\slashed{E}_{T}$ and $\text{jets} + \g$ plus $\slashed{E}_{T}$ searches at 13 TeV LHC with an increased integrated luminosity to $136$ fb$^{-1}$. Our findings demonstrate that the width of the GID significantly surpasses that of the KKPD in the case of $\delta=2$ and should be taken into account when $\delta=4,6$, especially in low-$\MD$ region. For $\delta=2$, pair-produced quarks and/or gluons decay through GID into a corresponding SM field and $\vecn$-th mode graviton, thereby resulting in $2$-jet plus missing transverse energy signal. In other cases, namely $\delta=4,6$, the GID may play a role in the last step of the decay cascade of the first level KK quarks and/or gluons. Hence, a final state signal comprising a photon from a decay $\g^{1} \to \g + \hat{G}^{\vecn}$ is more pronounced. It is possible that one of the decaying partons follows KKPD and the other decays through GID. In this case, jets $+ \g + \slashed{E}_{T}$ signal can be realized, where a photon originates from gravitationally decaying $\g^{1}$, and jets come from the decay of a parton following KK-number preserving decay. We scanned the fundamental parameters of the fat-mUED model, namely $\Rinv$, $\MD$ within ranges of $[1-3]$ TeV and $[5-15]$ TeV, correspondingly, for $\delta=2, 4$ and $6$ cases. We found that for $\delta=2(6)$, ATLAS $13$ TeV multi-jet ($\text{jets}+\g~$) $+ \slashed{E}_{T}$ searches excludes $\Rinv < 2.9(2.95)$ TeV for $\MD=15$ TeV. Furthermore, for $\delta=4$, $\Rinv < 2.95$ TeV at $M_{D}=15$ TeV and $\Rinv<2850$ GeV at $M_{D}=5$ TeV are excluded by the ATLAS multi-jet and $\text{jets}+\g$ searches, correspondingly.   

\acknowledgments
This work is supported in part by TUBITAK through project number 121F051. The numerical calculations reported in this paper were fully/partially performed at TUBITAK ULAKBIM, High Performance and Grid Computing Center (TRUBA resources). DK thanks to K. Ghosh for discussions. We also thank to Ismail Turan for permission to use his computing resources and for careful reading of the manuscript. 

\bibliographystyle{JHEP}
\bibliography{fatmued.bbl}

\end{document}